\documentclass[a4paper,11pt,dvips]{article}
cm \voffset = -2truecm

\usepackage{graphicx}
\usepackage{amsmath}
\usepackage{amssymb}
\usepackage{latexsym}
\usepackage{subcaption}
\usepackage[colorlinks]{hyperref}
\usepackage{color}
\usepackage{enumerate}
\usepackage{enumitem}
\usepackage{cite}
\usepackage{makeidx}
\usepackage{colortbl}

\newcommand{\bra}{\begin{array}}
\newcommand{\era}{\end{array}}
\newcommand{\beq}{\begin{equation}}
\newcommand{\eeq}{\end{equation}}
\newcommand{\beqar}{\begin{eqnarray}}
\newcommand{\eeqar}{\end{eqnarray}}

\def\BC{\bb C}
\def\_\BC{\bbi C}

\def\( {\left(}
   \def\) {\right)}
\def\[ {\left[}
\def\] {\right]}
\def\no2 {{\textstyle{n\over 2}}}

\def\dag {{\dagger}}
\begin{document}
\begin{titlepage}
\setcounter{page}{1}
\renewcommand{\thefootnote}{\fnsymbol{footnote}}

\begin{flushright}
%ucd-tpg 10-01\\
%arXiv:yymm.xxxx
\end{flushright}

\vspace{5mm}
\begin{center}
{\Large \bf {Energy Levels of Quantum Ring in ABA-Stacked Trilayer
Graphene  }}

\vspace{5mm}

{\bf Abdelhadi Belouad}$^{a}$, {\bf Youness Zahidi}$^{b}$, {\bf
Ahmed Jellal\footnote{\sf ajellal@ictp.it --
jellal.a@ucd.ac.ma}}$^{a,c}$, {\bf Abdelhadi Bahaoui}$^{a}$

\vspace{5mm}

{$^{a}$\em Theoretical Physics Group,  %Department of Physics,
Faculty of Sciences, Choua\"ib Doukkali University},\\
{\em PO Box 20, 24000 El Jadida, Morocco}

{$^{b}$\em MATIC, FPK, Hassan 1 University, Khouribga, Morocco}

{$^c$\em Saudi Center for Theoretical Physics, Dhahran, Saudi
Arabia}

\vspace{3cm}

\begin{abstract}
We present the solutions of the energy spectrum % expressions for the eigenstates and eigenvalues
of charge carriers confined in quantum ring in
ABA-stacked trilayer graphene subjected to a perpendicular
magnetic field. The calculations were performed in the context of
the continuum model by solving the Dirac equation for a zero width
ring geometry, i.e. freezing out the carrier radial motion. We show that the
obtained energy spectrum exhibits different symmetries with respect to the  magnetic
field and other parameters. The application of a potential shift the energy
spectrum vertically while the application of a
magnetic field breaks all symmetries. We compare our results with
those of
of the ideal quantum ring in
monolayer and bilayer graphene.

\end{abstract}
\end{center}
\vspace{3cm}
\noindent PACS numbers:  81.05.ue, 81.07.Ta, 73.22.Pr\\
\noindent Keywords: ABA-stacked trilayer graphene, quantum ring, Dirac equation, magnetic field, symmetries.
\end{titlepage}

%%%%%%%%%%%%%%%%%%%%%%%%%%%%%%%%%%%%%%%%%%%%%%%%%%%%%%%%%%%%
\section{Introduction}
%%%%%%%%%%%%%%%%%%%%%%%%%%%%%%%%%%%%%%%%%%%%%%%%%%%%%%%%%%%%

Graphene being one atom thick layer of carbon atoms ordered into a
honeycomb lattice, has attracted a lot of theoretical and
experimental research \cite{Geim07,Guinea09,Castro09,Martino11}.
This is due to its unconventional electronic properties and
promising applications to nanoelectronics. In fact, many
intriguing transport phenomena, such as anomalous quantum Hall
effect \cite{Novoselov05,Y.Zhang05} and  Klein tunneling
\cite{Katsnelson06} have been reported. In addition graphene has
an unique band structure, which is gapless and exhibits a linear
dispersion relation at two inequivalent points $K$ and $K'$ in the
Brillouin zone. This makes charge transport in graphene
substantially different from that of conventional two-dimensional
electronic systems. The equation describing the electronic
excitations in graphene is formally similar to the Dirac equation
for massless fermions, which travel at a speed of the order of
$v_F \approx 10^6$ m/s \cite{Semenoff84,DiVincenzo84}.

Few layers of graphene can be stacked above each other to form
what is called stacked graphene systems. There exist important stacking
configurations, that differently depending on the horizontal shift
of graphene planes. Every sequence of stacking behaves like a new
material, which leads to different electronic properties
\cite{A.Avetisyan10,M.F.Craciun09,J.H.Warner10}.
For example, for three coupled graphene sheets, it is known as
trilayer graphene. Each honeycomb contains three cells where each
cell consists of two carbon atoms named $A$ and $B$.
Interestingly, it can be formed in two major stacking types: ABA
(Bernal) and ABC (rhombohedral) stacking
\cite{K.F.Mak10,F.Zhang10}. Recently, trilayer graphene has
attracted much interest and have been proposed as promising
candidates for the future nanoelectronics. This is due to the fact
that its electronic structure is different from the bands found in
more studied monolayer and bilayer graphene.
Trilayer
graphene in the Bernal stacking (ABA) is a common hexagonal
structure which has been found in graphite. In trilayer graphene
there are three coupled layers in the bottom, middle and top
\cite{McCann10}. The $A2$ atom from middle layer is directly above
$B1$ atom from bottom layer and below $B3$ atom from top layer.
The energy bands of trilayer graphene are constituted of two
separate types
\cite{Lu06,Guinea06,Latil06,Partoens07,M.Koshino08,Aoki07}: two
linear bands who looks like the bands in monolayer graphene and
four parabolic bands similar to those in bilayer graphene.

It has been shown that graphene can be cut in many different
shapes and sizes giving rise a very important class of quantum
devices. This open the door to the fabrication of graphene
nanodevices through the experimental obtaining of graphene quantum
rings \cite{Schnez08}, quantum dots
\cite{Novoselov07,Ponomarenko08} and also antidot arrays
\cite{Schen08}. This mean that the electronic properties of
graphene are induced by its size and shape. In addition, the
confinement has an effect on the electronic structure of graphene
quantum dots. In fact, some theoretical studies were reported this
effect on the energy levels of graphene quantum dots with
different geometries, sizes and types of edge
\cite{Yamamoto06,Zhang08}. Moreover, the electronic structure of
graphene quantum rings \cite{Recher07,DSL} and of graphene antidot
lattices \cite{1Pedersen08,2Pedersen08} have also been
investigated. The graphene-based quantum rings have been
obtained experimentally by lithographic techniques
\cite{Huefner10}. Recently, these systems have been studied
theoretically in monolayer graphene \cite{Recher07}, AB-stacked
bilayer \cite{Zarenia10} and also in AA-stacked bilayer graphene
\cite{Zahidi17}.

In this paper we consider a quantum ring in ABA-stacked trilayer
graphene in the presence of an external magnetic field and a
%confinement
potential.
By freezing out the carrier radial motion (ideal quantum
ring) for zero and non zero magnetic field, we determine the solutions
of the energy spectrum
%in terms
as six band solutions.
Indeed, these bands are composed of the energy levels of a bilayer
\cite{Zarenia10} and those of a single layer graphene.
We show that our energy spectrum exhibits different symmetries
related to the magnetic field and other parameters.
We numerically study our results
to underline the behavior of the present
as well as compare them with
those for circular
ideal ring in monolayer graphene and bilayer graphene.

The present paper is organized as follows. In section $2$, we fix our
problem by setting the Hamiltonian describing a quantum ring in ABA-stacked trilayer
graphene in the presence of an external magnetic field and a
%confining
potential. Subsequently, we use
the eigenvalue equation to obtain the six band solutions in terms the magnetic field, radius of ring
and the interlayer coupling parameter.
% expressions for the
%energy spectrum.
In section $3$, we give different numerical results related to the energy spectrum where
the symmetries will be fixed under some transformations related to magnetic field
and a quantum number. We conclude our results and discussions in the final section.

%%%%%%%%%%%%%%%%%%%%%%%%%%%%%%%%%%%%%%%%%%%%%%%%%%%%%%%%%%%%%%%%%%%%%%%ù
\section{Problem setting}
%%%%%%%%%%%%%%%%%%%%%%%%%%%%%%%%%%%%%%%%%%%%%%%%%%%%%%%%%%%%%%%%%%%%%%%

\begin{figure}[!ht]
  \centering
  \includegraphics[width=9cm, height=6cm]{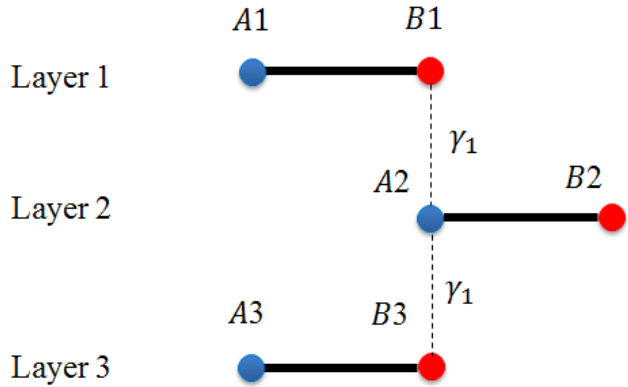}
  \caption{\sf{Crystallographic structures showing the relative position of
  sublattices $Ai$ and $Bi$ ($i=1,\ 2,\ 3$) for ABA-stacked trilayer
graphene with intra-layer $\gamma_1$ hopping amplitudes.}}
\label{cc}
\end{figure}

We consider a quantum ring in  ABA-stacked trilayer graphene  subjected to a magnetic field and
%confining
a potential. This
system consists of three coupled layers, each with carbon atoms
arranged in a honeycomb lattice, including pairs of inequivalent
sites ${A1,B1}$, ${A2,B2}$ and ${A3,B3}$ in the top, middle, and
bottom layers, respectively. The layers are arranged as shown in
Figure \ref{cc}, such that pairs of sites $B1$ and $A2$, and $A2$
and $B3$, lie directly above or below each other. The parameter
$\gamma_1=0.4 \ eV$ \cite{Dresselhaus02} describes strong
nearest-layer coupling between sites ($A2$-$B1$ and $A2$-$B3$)
that lie directly above or below each other.
%For the typical $\gamma_1=0.4 \ eV$ value for ABA-stacked trilayer graphene we
%quote.

In a basis with components of
$\Psi=(\Psi_{A1},\Psi_{B1},\Psi_{A2},\Psi_{B2},
\Psi_{A3},\Psi_{B3})^T$, the Hamiltonian of our
system can be written as
%The ABA-stacked trilayer Hamiltonian
%[7, 28, 40, 41]
%is
\begin{equation}
\label{eq1} H_{ABA} =
\begin{pmatrix}
u_1 & \pi^\dag & 0 & 0 & 0 & 0 \\
\pi & u_1 & \gamma_1 & 0 & 0 & 0 \\
0 & \gamma_1 &  u_2 & \pi^\dag & 0 & \gamma_1  \\
0 & 0 & \pi & u_2 & 0 & 0 \\
0 & 0 & 0 & 0 & u_3 & \pi^\dag \\
0 & 0 & \gamma_1 & 0 & \pi & u_3 \\
\end{pmatrix}
\end{equation}
where  $\Psi_{Ai}$ ($\Psi_{Bi}$) are the
envelope functions associated with the probability amplitudes of
the wave functions on the sublattice $A$ ($B$) of the
$i^{\texttt{th}}$ layer ($i = 1, 2, 3$),
$u_i$ ($i=1, 2, 3)$ is the potential in each layer, $v_F=10^6$ m/s
is the Fermi velocity.
 $\pi$ and $\pi^\dag$ are the momentum operators in polar
coordinates \beqar &&{\pi=v_Fe^{i\theta}\left[-i\hbar
\left(\frac{\partial}{\partial_r}+\frac{i\partial}{r\partial_\theta}\right)+i\frac{eB_0r}{2}\right]}\\
&&{\pi^\dag=v_Fe^{-i\theta}\left[-i\hbar
\left(\frac{\partial}{\partial_r}-\frac{i\partial}{r\partial_\theta}\right)-i\frac{eB_0r}{2}\right]}
\eeqar
in which the symmetric gauge is used to fix the
vector potential $A= \frac{B_0r}{2}$.

In the next we will present analytical expression of the
eigenstates and energy levels of ideal quantum ring with radius
$R$ created with ABA-stacked trilayer graphene. For this system,
%For an ideal ring with radius $R$,
the momentum of the charge carriers in the radial direction is
zero. By freezing out the carrier radial motion, the
four-component wave function becomes
\begin{equation}
\label{eq4} \Psi(r,\theta)=\begin{pmatrix}
\Phi_{A1}(r)e^{i(m-1)\theta} \\
i\Phi_{B1}(r)e^{im\theta}\\
i\Phi_{A2}(r)e^{im\theta}\\
\Phi_{B2}e^{i(m+1)\theta} \\
\Phi_{A3}(r)e^{i(m-1)\theta} \\
i\Phi_{B3}(r)e^{im\theta}
\end{pmatrix}
\end{equation}
where $m\in \mathbb{Z}$ is eigenvalues of the angular momentum label.
Now
solving the Dirac equation $H\Psi(R,\theta)=E\Psi(R,\theta)$
%and using the symmetric gauge $A=\left(0,\frac{B_0r}{2},0\right)$,
to obtain the following system of coupled differential equations
%The eigenstates of Eq. \ref{eq1} are six component spinors which, in polar coordinates, are given by
%we obtain the following system of coupled differential equations
\begin{eqnarray}
\label{eq5} %
%\begin{array}{llllll}
&&(u_1-\epsilon)\Phi_{A1}(R)+(m+\beta)\Phi_{B1}(R)=0  \\
&&(m-1+\beta)\Phi_{A1}(R)+(u_1-\epsilon)\Phi_{B1}(R)+\gamma\Phi_{A2}(R)=0\\
&&\gamma\Phi_{B1}(R)+(u_2-\epsilon)\Phi_{A2}(R)-(m+1+\beta)\Phi_{B2}(R)+\gamma\Phi_{B3}(R)=0\\
&&(m+\beta)\Phi_{A2}(R)-(u_2-\epsilon)\Phi_{B2}(R)=0 \\
&&(u_3-\epsilon)\Phi_{A3}(R)+(m+\beta)\Phi_{B3}(R)=0 \\
&&\gamma\Phi_{A2}(R)+(m-1+\beta)\Phi_{A3}(R)+(u_3-\epsilon)\Phi_{B3}(R)=0
%\end{array}%
\end{eqnarray}
where we have set the dimensionless quantities $E_0=\frac{\hbar
v_F}{R}$, $\gamma=\frac{\gamma_1 }{E_0} $, $\epsilon=\frac{E
}{E_0}$, $\beta =\left(\frac{eB_0 }{2\hbar}\right)R^2$ and
$U_i=\frac{u_i}{E_0}$ ($i=1, 2, 3$). After some straightforward
algebra and fixing the potential as $u_i=u$, we derive the six energy band solutions
\begin{eqnarray}
&& {S_{1}^{s}=s\sqrt{(m+\beta)(m+\beta+1)}}\label{e1q6}\\
 &&
{(S_{2}^{s})^2=(m+\beta)^2+(\gamma')^2/2+s\sqrt{(m+\beta)^2(1+(\gamma')^2)+(\gamma')^4/4}}\label{e1q7}
\end{eqnarray}
where $S=\epsilon-U$, $s=\pm 1$, $\gamma'=\sqrt{2}\gamma$. We notice that \eqref{e1q6} corresponds to
energy levels of a single layer graphene, whereas
\eqref{e1q7} coincides (apart from a numerical factor
$\sqrt{2}$ in front of $\gamma$) with energy levels of a bilayer
graphene \cite{Zarenia10}. This is expected since
the low energy band structure of ABA-staked trilayer graphene for
zero magnetic field consists of two single layer graphene-like
bands and four AB-staked bilayer graphene-like bands
\cite{S.Yuan11,Van.Duppen13}.
%as it has been discussed above
%%%%%%%%%%%%%%%%%%%%%%%%%%%%%%%%%%%%%%%%
%Eq.\ref{eq6} can also be written as
%\begin{equation}\label{eq8}
%\epsilon=\pm\sqrt{(m-m_-)(m-m_+)}
%\end{equation}
%where
%\end{equation}
From \eqref{e1q6}, we see that the energy spectrum for an ideal quantum ring,
is real when $m < -\beta - 1$ or  $m > -\beta$ and imaginary
otherwise.
%%%%%%%%%%%%%%%%%%%%%%%%%%%%%%%%%%%%%%%%
However for \eqref{e1q7}, the four solutions for the energy %, these
are real
only when $|m+\beta|\geq1$. In the opposite case of $|m+\beta|< 1$ (or
equivalently $-1+\beta < m < 1 - \beta$) we have $S_2^{-} < 0$ and
consequently the corresponding energies are imaginary. To solve
this problem we propose the limit of $\gamma\gg m+\beta$, we obtain
\beq
(S_{2}^{-})^2=\frac{1}{(\gamma')^2}(m+\beta)^2((m+\beta)^2-1)
\eeq
and thus the  energy solutions are given by
\begin{equation}\label{eq8}
(S_{2}^{-})\simeq\pm\frac{1}{\gamma'}\sqrt{(m+\beta)^2((m+\beta)^2-1)}.
\end{equation}
%We replace $S$ by $\epsilon-U$ in
From \eqref{e1q6} and \eqref{e1q7} we end up with the six band solutions
%\begin{eqnarray}
\begin{eqnarray}
&& E_{1}^{s}=\frac{\hbar v_F}{R}\left[U+s\sqrt{(m+\beta)(m+\beta+1)}\right]\label{eq9}\\
&& E_{2}^{+}=\frac{\hbar v_F}{R}\left[U\pm\sqrt{(m+\beta)^2+(\gamma')^2/2+\sqrt{(m+\beta)^2(1+(\gamma')^2)+(\gamma')^4/4}}\right]\label{eq10}\\
&& E_{2}^{-}=\frac{\hbar
v_F}{R}\left[U\pm\frac{1}{\gamma'}\sqrt{(m+\beta)^2((m+\beta)^2-1)}\right]\label{eq11}.
\end{eqnarray}
%\end{eqnarray}
These will numerically be analyzed based on different consideration of
the involved parameters, which concern the external magnetic field
and the eigenvalues of the angular momentum.

%%%%%%%%%%%%%%%%%%%%%%%%%%%%%%%%%%%%%%%%%%%%%%%%%%%%%%%%%%%%%%%%%
\section{Results and discussions}
%%%%%%%%%%%%%%%%%%%%%%%%%%%%%%%%%%%%%%%%%%%%%%%%%%%%%%%%%%%%%%

\begin{figure}[!ht]
  \centering
  \includegraphics[width=8.5cm, height=8cm]{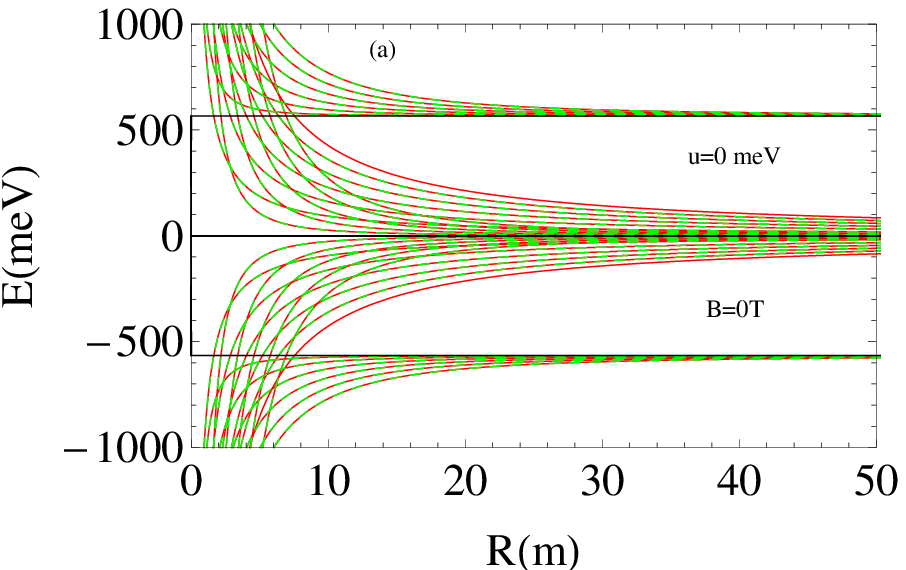}\includegraphics[width=8.5cm, height=8cm]{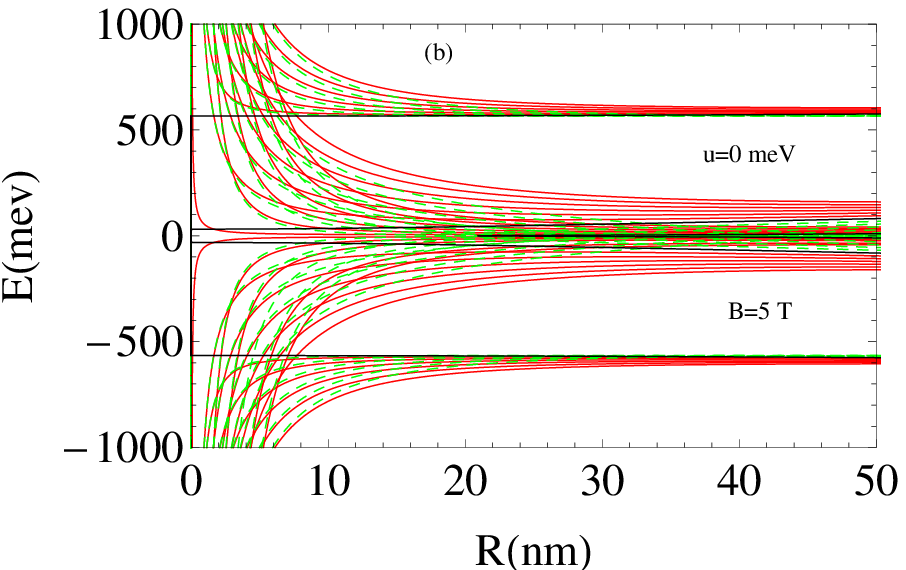}\\
   \includegraphics[width=8.5cm, height=8cm]{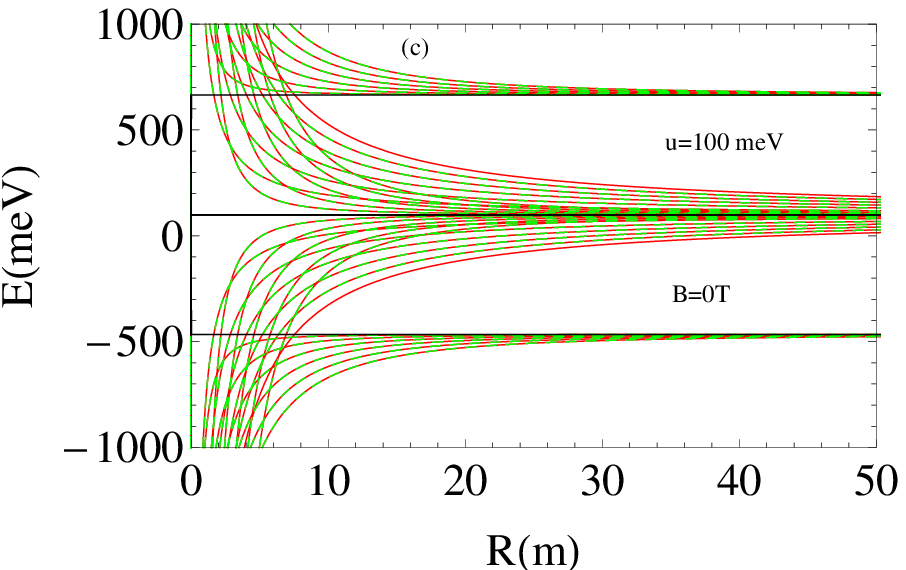}\includegraphics[width=8.5cm, height=8cm]{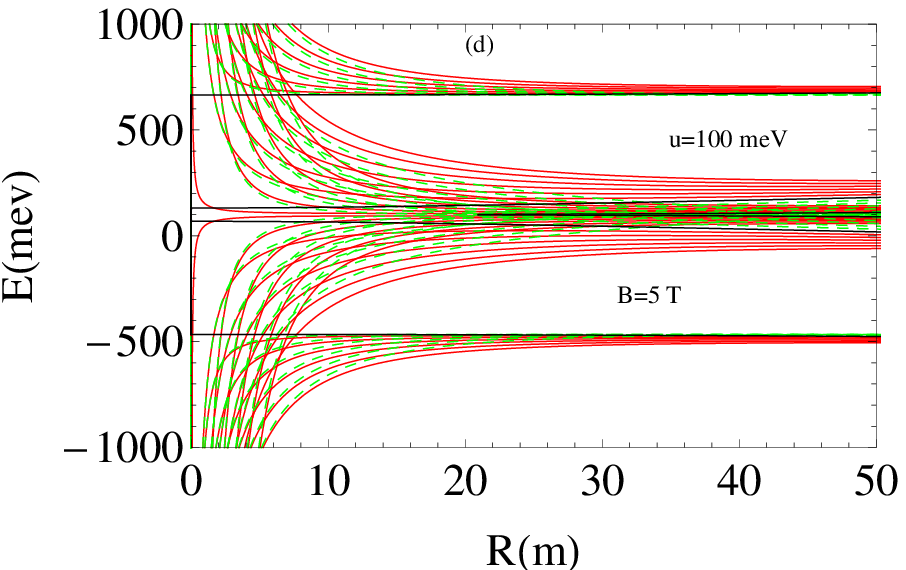}\\
\caption{\sf{Energy levels of a trilayer graphene quantum ring as
function of ring radius $R$ with $B=0$ T ((a): $u=0$ meV and (c): $u=100$ meV)
and $B=5$ T ((b): $u=0$ meV and (d): $u=100$
meV). The total angular quantum number $-6\leq m\leq-1$ (green),
$1\leq m \leq 6$ (red)
   and $m=0$ (black).}} \label{f1}
  \end{figure}

In Figure \ref{f1}, we plot the energy levels of an ideal quantum
ring in ABA-stacked trilayer graphene as a function of the ring
radius $R$ with $B=0$ T ((a) $u=0$ meV, (c) $u=100$ meV) and $B=5$
T ((b) $u=0$ meV, (d) $u=100$ meV). The green and red curves
correspond, respectively, to $-6\leq m\leq-1$ and $1\leq m \leq 6$
while the black one correspond to $m=0$.
%The total angular
%quantum number $-6\leq m\leq-1$ (Green, dashed), $1\leq m \leq 6$
%(Red) and $m=0$ (Black).
For zero magnetic field (Figure \ref{f1}(a) and (c)), we can see
that the energy spectrum shows two sets of levels. One similar to
the monolayer  and the other corresponding to the AB-stacked
bilayer graphene \cite{Zarenia10}.
% In fact, the two low energy band structure of
%ABA-staked trilayer consists of two single layer-like bands and
%four bilayer-like bands \cite{}.

For a deep understanding of our results,
we distinguish two cases. The first is
%We notice that for
zero magnetic field where the the energy branches take the
following form
\begin{equation}\label{eq12}
E_{1}^{s}=u+s\frac{\hbar v_F}{R}\sqrt{m(m+1)}
\end{equation}
which
%In this case
have a $1/R$ dependence, however
for large $R$, the set of levels converge to the potential $u$.
% however the
%other set of levels converge to $\sqrt{2}\gamma$.
Note that for $m=0$ and $m=-1$, the energy $E_{1}^{s}=u$ is
independent of $R$. The second is nonzero magnetic field where the right panel
(Figure \ref{f1}(b) and (d)) shows that the branches have an
approximately linear dependence on the ring radius for large $R$. This is
%In particular we have
\beq E_{1}^{s}=u+s\frac{v_FeB}{2}R. \eeq However, for small $R$,
all branches diverge as $1/R$ except for two values $m=0$ and
$m=-1$, which give the result \beq
E_{1}^{s}=u+s\frac{\hbar}{R}\sqrt{m(m+1)} \eeq
and in addition, for zero magnetic field and for large radius the
energy branches approach as $E\longrightarrow \pm \sqrt{2}\gamma$.
%by evaluation \eqref{e1q7}, we can clearly show that the
%second set of energy levels are shifted up by by
%$\sqrt{2}\gamma_1$.
We notice that from \eqref{e1q7} and for zero magnetic field, we
have the energy branches
\begin{eqnarray}
&&E_{2}^{+}=u\pm\sqrt{\frac{(\hbar
v_F)^2}{R^2}m^2+\gamma_{1}^{2}+\sqrt{\gamma_{1}^{4}+\left(m^2\frac{(\hbar
v_F)^4}{R^4}+2\gamma_{1}^{2}\frac{(\hbar
v_F)^2}{R^2}\right)}}\label{eq14}\\ && E_{2}^{-}=u\pm\frac{(\hbar
v_F)^2}{\sqrt{2}\gamma_1 R^2}\sqrt{m^2(m^2+1)} \label{eq13}.
\end{eqnarray}
From these results, one can deduce different conclusions. Indeed,
Firstly, we can clearly show that, for zero magnetic field, the second set
of energy levels depends on $1/R^2$. Secondly,
%%%%%%%%%%%%%%%%%%%%%%%%%%%%%%%%%%%%%%%%%%%%%%%%%%%%%%%Ã¹
in the limit  $R\longrightarrow \infty$, we have
$E^{-}_2\longrightarrow u$ and $E^{+}_2\longrightarrow
u+\sqrt{2}\gamma_1$. Thirdly when $R\longrightarrow 0$, the
behavior of the spectrum is different and the corresponding energy
levels diverge.
%$E_{2}^{-}=U+\pm\frac{(\hbar v_F)^2}{\sqrt{2}\gamma R^2}\sqrt{(m^2(m^2+1)}$ and $E_{2}^{+}=U+\sqrt{\frac{(\hbar v_F)^2}{\sqrt{2}\gamma R^2}\sqrt{(m^2(m^2+1)}}$
In addition, we note that the applied potential has different
effect on the system behavior compared to that in the case of AB-stacked
bilayer quantum ring \cite{Zarenia10}. Indeed, in our system when
$u=100$ meV all sets are shifted up by $u$, however in the case of
ideal quantum rings in AB-stacked bilayer the application of the
potential open a gap in the energy spectrum.

Now we return back to investigate the basic features of \eqref{eq10} and \eqref{eq11}. Indeed,
in the case of non zero magnetic field and for small ring radius
$R$, both equations reduce %\eqref{eq10} and \eqref{eq11} becomes
\begin{eqnarray}
 && E_{2}^{+}=u\pm\frac{\hbar
v_F}{R}\sqrt{m^2+\mid m\mid} \label{eq16}\\
&& E_{2}^{-}=u\pm\frac{(\hbar v_F)^2}{\sqrt{2}\gamma_1
R^2}\sqrt{m^2(m^2+1)}\label{eq15}
\end{eqnarray}
depending on
%From \eqref{eq16}, we can show that
%the spectrum behaves as
$1/R$
%dependence, however from \eqref{eq16} we show that the spectrum
%depends on
and $1/R^2$, respectively. However for large radius $R$, \eqref{eq10} and
\eqref{eq11} give the results
%reliance to
\begin{eqnarray}\label{eq18}
&&E_{2}^{+}=u\pm\frac{(v_FeB/2)^2}{\sqrt{2}\gamma_1 }R^2
\\  && E_{2}^{-}=u\pm\frac{eBv^{2}_{F}}{2} R. \label{eq17}
\end{eqnarray}
which  explain the approximately $R^2$ dependence of the energy
branches on the radius of the ring and  also %this can clearly explain
the approximately linear dependence of the energy branches on the
ring radius. It is important to note that, for zero magnetic
field, all branches are twofold degenerate. However, the
application of a nonzero magnetic field breaks the degeneracy of
all branches.

\begin{figure}[!ht]
  \centering
  \includegraphics[width=8.5cm, height=8cm]{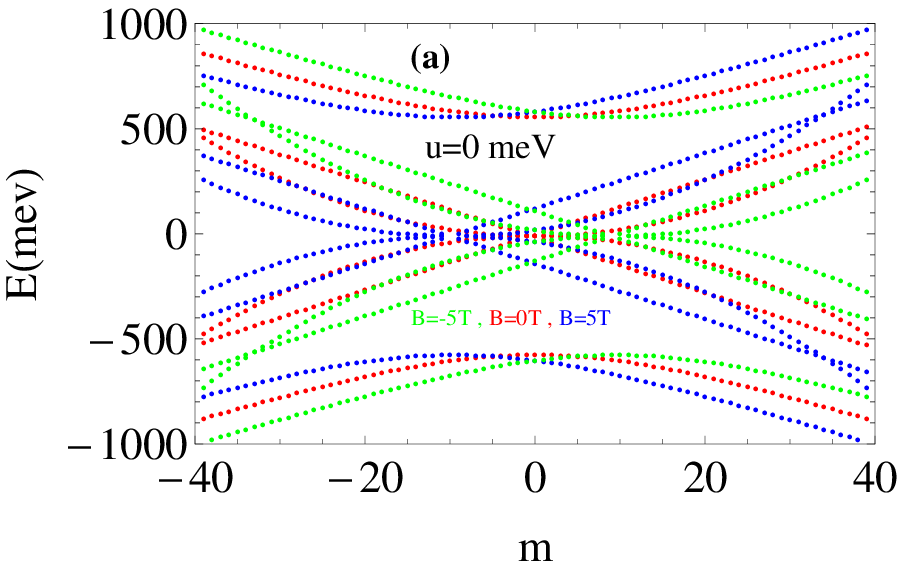}\includegraphics[width=8.5cm, height=8cm]{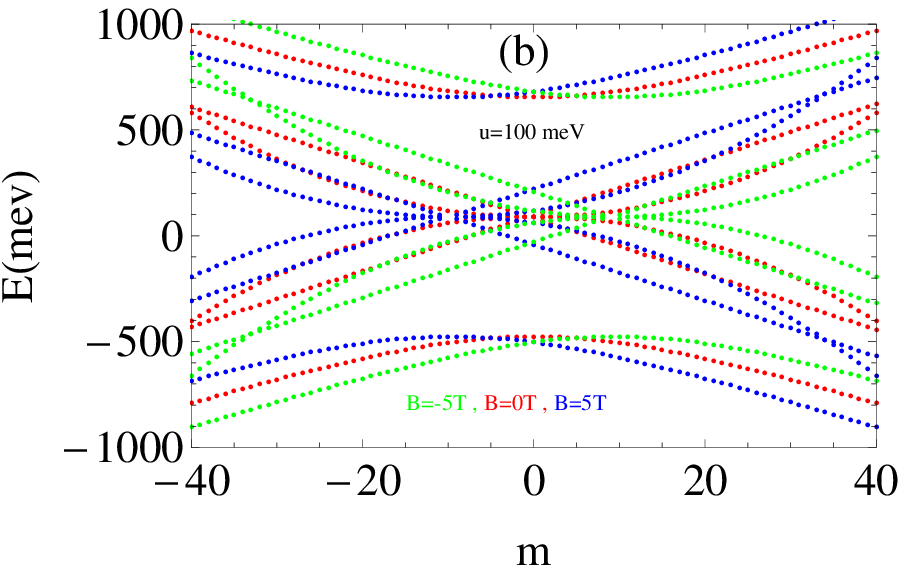}\\
  \caption{\sf{Energy levels of a trilayer graphene quantum ring as as function of total angular momentum label $m$
  for $B=-5$ T (geen), $B=0$ T (reed) and $B=5$ T (blue) and $R=50$ nm. (a): $u=0$ meV, (b): $u=100$ meV.}} \label{f2}
\end{figure}

In Figure \ref{f2}, we plot the energy levels of an ideal quantum
ring of ABA-stacked trilayer graphene as a function of the angular
momentum $m$ for three different values of the magnetic field ($B=
-5$ T) (green), $B=0$ T (red) and $B=5$ T (blue) for $R=50$ nm
with $u=0$ meV (Figure \ref{f2}(a)) and $u= 100$ meV (Figure
\ref{f2}(b)). The energy spectrum represents two bands dispersed
with an energy \eqref{e1q6}, which is quite similar to that of an
single layer (linear dispersion), and four bands of type bilayer
graphene (parabolic dispersion) \eqref{eq8} and \eqref{eq9}. We
notice that, like the case of monolayer \cite{Zarenia10} and
AB-stacked bilayer graphene \cite{daCosta14}, the electron energy
present a minimum for a particular value of the angular momentum
$m$. The energy minimum for $B=-5$ T is given by $m=10$. However,
the energy minimum for $B=0$ T and $B=5$ T, are respectively,
given by $m=0$ and $m=10$. This can be explained by the fact that
from \eqref{eq9}-\eqref{eq11}, the spectra are invariant under the
transformation $B\longrightarrow -B$, $m\longrightarrow -m$ and
$m\longrightarrow -(m + 1)$. Therefore, the energies satisfy the following
%related by the
symmetries with respect to the above transformations
%relations
\begin{eqnarray}\label{eq18}
&&E_{1}^{s}(m,B)=E_{1}^{s}(-m-1,-B)\\
&&E_{2}^{+}(m,B)=E_{2}^{+}(-m,-B)\\
&&E_{2}^{-}(m,B)=E_{2}^{-}(-m,-B).
\end{eqnarray}
These symmetries are also
%which are also
present in the case of ideal quantum ring in
monolayer and bilayer graphene systems \cite{Zarenia10}.

\begin{figure}[!ht]
  \centering
  \includegraphics[width=8.5cm, height=8cm]{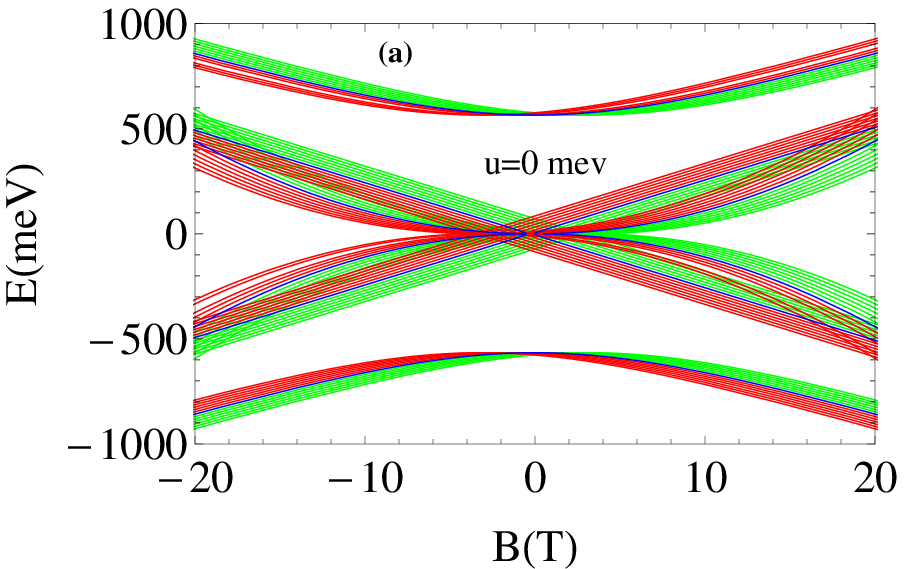}\includegraphics[width=8.5cm, height=8cm]{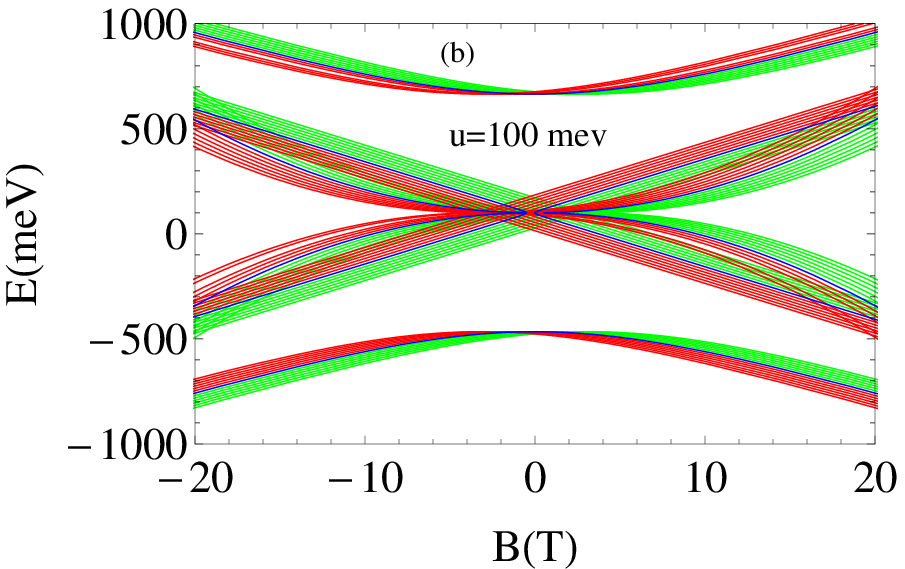}\\
  \caption{\sf{Electron and hole energy states of a trilayer graphene quantum ring as as function
  of external magnetic field $B$ for $R=50$ nm. $-6\leq m\leq-1$ (green curves),
  $1\leq m \leq 6$ (red curves) and $m=0$ (blue curves).}} \label{f3}
\end{figure}

The energy levels of the electrons and holes for an ideal quantum
ring in a ABA-stacked trilayer graphene as function of the
external magnetic field are shown in Figure \ref{f3} for $R=50$
nm, with (a): $u=0$ meV, (b): $u=100$ meV. The energy levels are
show for the quantum number $-6\leq m\leq-1$ (green), $1\leq m
\leq 6$ (red) and $m=0$ (blue).
%In Figure \ref{f3}(a), we plot the energy levels of the electrons
%and holes for the ideal quantum ring in a ABA-stacked trilayer
%graphene for $u=0$ meV, while in Figure \ref{f3}(b) let us plot
%the energy levels of electrons and holes for the ideal quantum
%ring in a stacked two-layered ABA graphene for $u\neq 0$.
We can clearly show, for $u=0$ meV (Figure \ref{f3}(a)), that the
energy spectrum presents two sets of energy levels, one is similar
to the energy levels of AB-stacked bilayer graphene quantum ring
that are parabolically disperses with $B$, and the second one
correspond to the monolayer graphene quantum ring that are
dispersing linearly with $B$. Note that when $u\neq0$, the energy
levels are shifted vertically by $u$ (Figure \ref{f3}(b)). This
results are not similar to that obtained for AA-stacked
\cite{Zahidi17} and AB-stacked \cite{Zarenia10} bilayer graphene
quantum ring, where the application of a potential $\neq 0$ open a
gap in the energy spectrum. From Figure \ref{f3}(b), we can show
that there is an asymmetry between the electron and hole states
caused by the application of the potential. Also, these results
show that the electron and hole energies are related by
\begin{eqnarray}\label{eq19}
&&E_{1(e)}^{s}(m,B)=-E_{1(h)}^{s}(-m-1,-B)\\
&&E_{2(e)}^{+}(m,B)=-E_{2(h)}^{+}(-m,-B)\\
&&E_{2(e)}^{-}(m,B)=-E_{2(h)}^{-}(-m,-B)
\end{eqnarray}
 where the indices $h(e)$ refer to holes (electrons). These results are similar
 to that obtained for a finite width quantum ring in monolayer graphene and
 AB-stacked bilayer graphene \cite{Zarenia10}. %, where  the electron and hole states are asymmetric.

 %%%%%%%%%%%%%%%%%%%%%%%%%%%%%%%%%%%%%%%%%%%%%%%%%%
\section{Conclusion}
%%%%%%%%%%%%%%%%%%%%%%%%%%%%%%%%%%%%%%%%

In summary, we have investigated the behavior of charge carriers
in ideal quantum rings of ABA-stacked subjected to a perpendicular
magnetic field.
% as function of the ring radius, the potential and
%the magnetic field.
The calculation was performed by solving the Dirac equation for a
zero width ring geometry (ideal ring). In the case of an ideal
ring with radius $R$, the momentum of the carriers in the radial
direction is zero, then we have to treat the radial parts of the
spinors as a constant.
%Our approach leads to
%analytical expressions for the energy spectrum in the continuum
%model, solving the Dirac equation for a zero-width ring geometry
%by freezing the radial support motion.
Our approach lead to analytical expressions of the energy
spectrum as a function of the ring radius and the magnetic field.

Our results show that the energy spectrum of an ideal quantum ring
presents two sets of states as a function of the ring radius $R$.
One corresponding to the monolayer graphene and the other
corresponding to the AB-stacked bilayer graphene quantum ring.
%The bands structure are of two separate types : two almost-linear
%bands reminiscent of the bands in monolayer-like graphene and four
%parabolic bands similar to those in AB-stacked bilayer-like.
%This raises the expectation that the electronic
%behavior will display no new features as compared to or bilayer
%graphene
%Our numerical results showed that the low energy band structure of ABA staked trilayer graphene consists of two single layer bands and four bilayer graphene bands.
We have shown that, for large $R$, the set of energy levels that
correspond to the monolayer graphene converge to $\sqrt{2}\gamma$.
The energy branches corresponding to the monolayer graphene have a
$1/R$ dependence, whereas the energy branches corresponding to the
bilayer graphene have a dependence $1/R^2$. This implied that the
energy spectrum converge to $u$ and $u + \sqrt{2}\gamma$ for a
very large radius and diverges when the radius tends to zero. In
the absence of the magnetic field, the energy levels are twofold
degenerate. The application of nonzero magnetic field broken all
the degeneracy of all branches and the obtained spectrum exhibited
different symmetries. % and has a dependence on the magnetic field.
In addition, our numerical results showed that the electron and
hole energy levels are not invariant under the transformation
$B\longrightarrow-B$, $m\longrightarrow-m$ and $m\longrightarrow-(m + 1)$. The application of a potential lead to a vertically
shifting of the energy spectrum. %by $u$.
These results are not
similar to those obtained for AA-stacked and AB-stacked bilayer
quantum ring, where the application of a potential open a gap in
the energy spectrum.

%%%%%%%%%%%%%%%%%%%%%%%%%%%%%%%
\section*{Acknowledgment}
%%%%%%%%%%%%%%%%%%%%%%%%%%%%%%

The generous support provided by the Saudi Center for Theoretical
Physics (SCTP) is highly appreciated by all authors.

\end{document}